\documentclass[conference]{IEEEtran}
\IEEEoverridecommandlockouts
\usepackage[dvipsnames,table]{xcolor}
\usepackage{cite}
\usepackage{amsmath,amssymb,amsfonts}
\usepackage{algorithmic}
\usepackage{graphicx}
\usepackage{textcomp}
\usepackage{svg}
\usepackage{subcaption}
\usepackage{tikz}
\usetikzlibrary{positioning, shapes.geometric, arrows.meta}
\usepackage{multirow}
\begin{document}
\title{Precoder Implementation and Optimization in 5G NR Massive MIMO Radio}
\author{\IEEEauthorblockN{
 Kalyani Bhukya \IEEEauthorrefmark{1}, Shahid Aamir Sheikh\IEEEauthorrefmark{2}, Radha Krishna Ganti\IEEEauthorrefmark{3}
 \IEEEauthorblockA{
 Department of Electrical Engineering\\
 Indian Institute of Technology
 Madras\\
 Email: \IEEEauthorrefmark{1}ee21s124@smail.iitm.ac.in,
         \IEEEauthorrefmark{2}shahidaamir100@tamu.edu,
 		\IEEEauthorrefmark{3}rganti@ee.iitm.ac.in
   }
   }
   }
\maketitle

\begin{abstract}
The evolution of 5G New Radio (NR) has brought significant improvements in signal strength and service
quality for users. By the integration of Multiple Input Multiple Output(MIMO) systems in communications, multiple data streams can be transmitted simultaneously across multiple antennas. Additionally, the incorporation of precoding in MIMO systems enables enhanced data rates and spectral efficiency. In the case of wireless networks,
precoders are used to steer high-gain beams that are intended for specific users.
The paper focuses on the implementation of $\textbf{16}$, $\textbf{32}$ and $\textbf{64}$ channel linear precoders in the Remote
Radio Head (RRH) of the indigenously developed 5G testbed at IIT Madras. These precoders have a memory
module which stores the channel matrices and a multiplier module which performs matrix multiplications between the channel matrices and user data within a slot duration of $\boldsymbol{500 \, \text{microseconds}}$. The system demonstrates DSP utilization levels of $\boldsymbol{9.75\%}$, $\boldsymbol{19.5\%}$, and $\boldsymbol{39\%}$ for $\boldsymbol{(16\times8)}$, $\mathbf{(32\times8)}$, and $\boldsymbol{(64\times8)}$ antenna-layer configurations respectively, while maintaining the BRAM (Block RAM) usage within $\boldsymbol{2.28\%}$, $\boldsymbol{3.91\%}$, and $\boldsymbol{7.16\%}$. Additionally, a throughput of $\boldsymbol{1.2 \, \text{Gbps}}$ with four active layers highlights the system's optimized performance under hardware constraints.

\end{abstract}


\section{Introduction}
%\begin{figure}[h!]
   % \centering
       % \includegraphics[clip, trim=0.2cm 11cm 0.5cm 7cm, width=0.50\textwidth]{system_model-3.pdf}
    %    \includegraphics[clip, trim=0.2cm 0cm 0.5cm 0cm, %width=0.50\textwidth]{5g_BBU.pdf}
  %  \caption{5G Base Station Architecture}
  %  \label{fig: 5G Base Station Architecture}
    %\caption{State diagram}
    %\label{fig:State diagram}
%\end{figure}
The emergence of 5G New Radio (NR) technology aims to address the increasing demand for higher data rates, enhanced reliability, and ultra-low latency communication \cite{5gintro8258595}. A key enabler of these capabilities is the adoption of massive MIMO (Multiple Input Multiple Output) systems, which employ a large number of antennas at the base station (BS) to serve multiple users simultaneously. To fully leverage massive MIMO, the implementation of precoders is essential, as they shape transmitted signals to improve signal quality and reduce interference for the users. This makes precoding a crucial factor in achieving high throughput and effective beamforming.

Efficient precoding requires complex matrix multiplications under strict timing constraints, which typically lies within a $500\mu s$ slot duration\cite{jeeva10485845}. Due to the heavy computational demand of traditional complex multipliers, Karatsuba's technique is utilized to reduce the load \cite{karat7375548}. FPGA-based implementations, particularly for massive MIMO systems, aim to optimize speed, resource usage, and power efficiency \cite{FPGA9355138}. Several studies highlights the importance of pipelining, fixed-point arithmetic, and hardware-optimized algorithms to meet performance requirements \cite{nitin5710875}. Despite the advances, balancing low-latency processing with resource efficiency remains challenging for real-time applications. This paper focuses on optimizing FPGA-based linear precoding for massive MIMO systems.

The current precoder implementation and optimization for massive MIMO configurations, with antenna-layer pairs represented as ($N_T$, $N_L$) = (16x8), (32x8), and (64x8), where $N_T$ indicates the number of transmit antennas and $N_L$ is the number of layers. By evaluating matrix storage and multiplication techniques, and by efficient utilization of DSP resources, it has been demonstrated that real-time processing can be achieved at frequencies of 250 MHz for multipliers and 312.5 MHz for memory access. 

\begin{figure}[h!]
    \centering
        \includegraphics[clip, trim=0.2cm 0cm 0.5cm 0cm, width=0.50\textwidth]{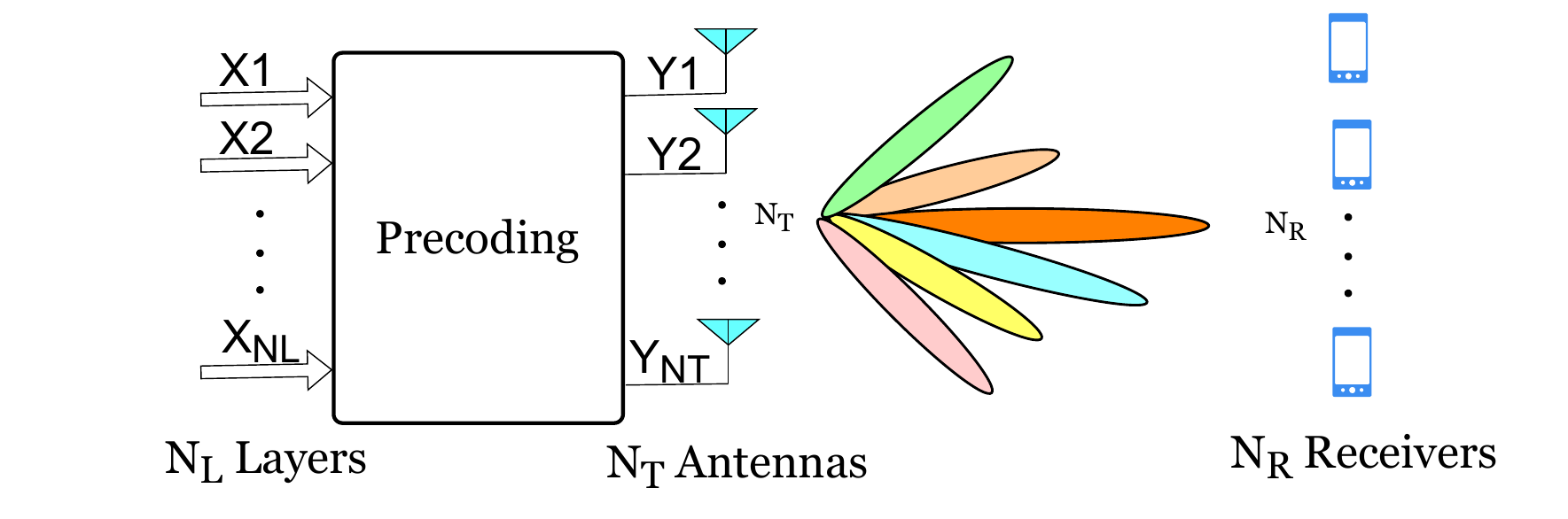}
    \caption{Precoding/Beamforming}
    \label{fig: Precoding/Beamforming}
    %\caption{State diagram}
    %\label{fig:State diagram}
\end{figure}

Fig. 1 illustrates the precoding and beamforming process, where multiple antennas direct the signals to various users.
% , optimizing signal quality and minimizing interference through adaptive beam adjustments.
The implementation discussed in this paper supports adaptive beamforming, enabling both narrow and wide beams to suit diverse user requirements in 5G networks.

The current paper is organized as follows: Section II provides the architecture required to build the downlink precoder. Section III describes the data flow taking place in the precoder. Section IV gives a detailed explanation of the testing and evaluation methods adopted in the current work. Section V reports the performance of the precoder design and discusses its effect on hardware utilization. Section VI provides the future directions that can be taken in precoder development in 5G NR. 

\section{Architecture Overview}

%\begin{figure}[ht]
%    \centering
%    \includegraphics[width=0.5\textwidth]{system_model.png}
%    \caption{Downlink Precoder Architecture: data processing from Precoder through Over-the-Air transmission}
 %   \label{fig: Precoder Architecture}
%\end{figure} 

\begin{figure*}[h!]
    \centering
        \includegraphics[clip, trim=0.2cm 0cm 0.5cm 0cm, width=0.9\textwidth]{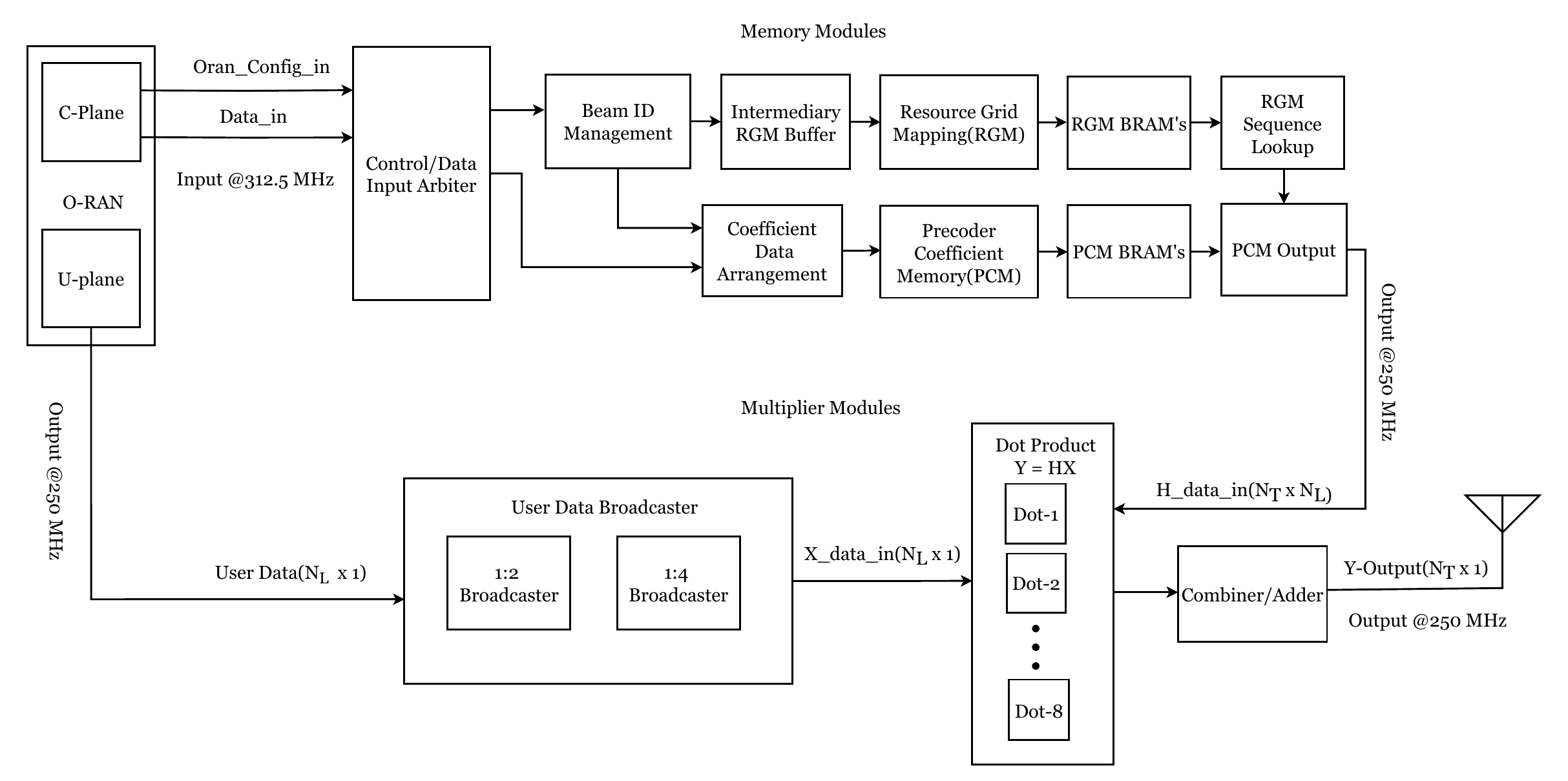}
    \caption{Downlink Precoder Architecture: data processing and transmission flow from O-RAN to precoder}
    \label{fig: Precoder Architecture}
    %\caption{State diagram}
    %\label{fig:State diagram}
\end{figure*}
% Explain where the precode lies, between which two modules in a 5G system. Then give an overall picture of the precoder design and purpose (input it takes and output it throws). 
In a 5G system\cite{5gtbiitm}, the downlink data flows from the Baseband Unit (BBU) to the Radio Remote Head (RRH) through the ORAN architecture\cite{ORAN_spec}. The BBU processes the control and user plane data before passing it to the precoder, which receives configuration and data information. The precoder then generates the beamfoming data, which is then forwarded to the digital front-end module for further processing, such as IFFT and guard band addition \cite{jeeva10485845}. The precoder consists of two major entities, namely, the memory module and matrix multiplier module operate sequentially to achieve this functionality. The detailed explanation of these  modules is in the subsequent subsections.
\subsection{Open-Radio Access Network (O-RAN)}
The O-RAN module serves as the primary interface for providing inputs to the precoder, through the two major planes: the control plane (C-plane) and the user plane (U-plane).
\begin{enumerate}
    \item Control plane (C-plane): The C-plane conveys control information to the precoder. This includes resource allocation, scheduling parameters, and beam configurations. It helps to map beamforming coefficients to the corresponding antenna ports using beam ID. 
    \item User plane (U-plane): The U-plane carries user data as IQ samples, which are mapped to resource elements based on the number of layers. This mapping enables the precoder to allocate data to the appropriate layers and antenna ports for efficient transmission. 
\end{enumerate}
\subsection{Memory module } The precoder memory module in the RRH is responsible for scheduling and storing precoding matrices for upcoming transmission slots. Leveraging a two-slot look-ahead structure, as highlighted in \cite{jeeva10485845}, it ensures the availability of real-time data. The module supports up to 64 user precoder matrices per slot and incorporates a ping-pong architecture, enabling simultaneous read and write operations. The clock domain crossings are efficiently managed using the XPM-CDC macros, ensuring reliable data transfer between the input and output domains.The various components within the memory subsystem (See Fig.\ref{fig: Precoder Architecture}) and their functions are as follows: 
\begin{enumerate}
    \item Control/Data input arbiter: This stage is responsible for accurately loading control and data packets into their respective pipelines in the correct order. The control packet includes input parameters such as slot\textunderscore ID, start\textunderscore symbol, num\textunderscore symbol, start\textunderscore PRB, num\textunderscore PRB, beam\textunderscore ID, Bundle\textunderscore PRB and Re\textunderscore Mask. 
    %Based on the channel configuration, the O-RAN C-plane sends a specific combination of control and data packets: [$1$ control $+$ $4$ data] packets for a 16-channel system, [$1$ control $+$ $8$ data] packets for a $32$-channel system, and [$1$ control $+$ $16$ data] packets for a $64$-channel system.
  %  Table \ref{tab:packet_formats} shows the number of bits required for each parameter in Oran\_config\_in and Mult\_config.
    
    \item Beam ID roster management: This stage is responsible for maintaining a ledger of all beam IDs encountered within the current slot and assigning each one with a unique Internal Address Number (IAN). The beam ID is a 16-bit unique identifier that contains beamforming information as specified in the 3\textsuperscript{rd} Generation Partnership Project (3GPP) standards\cite{3gpp_38_211}. 
    % Of these 16 bits, only 7 bits used for tagging coefficient matrices. 
    % The IAN is compared against stored beam IDs to ensure that the data is processed and stored with the correct precoding matrix.
      
    \item Intermediate resource grid mapping (RGM) buffer: RGM is a variable cycle operation, hence a buffer is inserted here so that the output of beam ID roster management can be stored and processed in sequence.

    \item Resource Grid Mapping (RGM) input: The RGM module retrieves scheduling information from the RGM buffer. It fills the resource grid with IANs (beam IDs) and sets a valid bit for successful writes. For unallocated PRBs, the valid bit remains unset, signaling the output to send zeros to conserve bandwidth. A ping-pong buffering scheme clears the RG before each write to support a two-slot look-ahead structure. Each RGM is $273 \times 14 \times (7 + 1) \; = \; 30,576$ bits, allowing up to $4$ IANs per cycle with a $32$-bit input. To handle a $1024$-cycle write overhead, two memory banks with $4$ BRAMs each can be used as shown in the  Fig.\ref{fig:RGM Ping Pong buffering}.
     \begin{figure}[htbp]
            \centering
                \includegraphics[clip, trim=0.0cm 0cm 0.6cm 0.0cm,width=0.40 \textwidth]{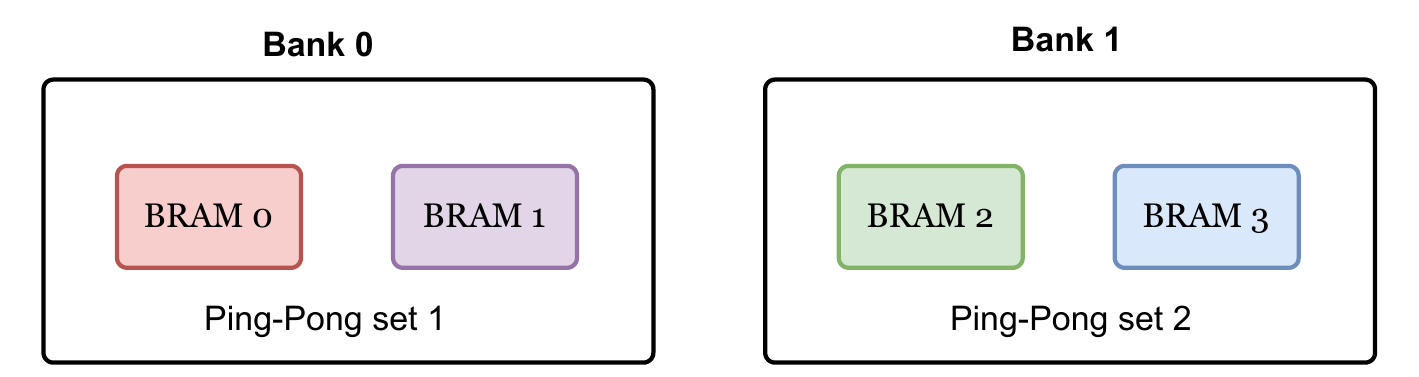}
            \caption{RGM Ping Pong buffering}
            \label{fig:RGM ping pong buffering}
       \end{figure}    
\begin{table}[h!]
    \centering
        \caption{RGM-BRAM operations with color coding}
    \renewcommand{\arraystretch}{1.0}
    \setlength{\tabcolsep}{2pt} % Adjust column padding
    \begin{tabular}{|p{1.0cm}|p{1.5cm}|p{1.0cm}|p{1.0cm}|p{1.0cm}|p{1.0cm}|}
        \hline
        & &  \textbf{2'b00} & \textbf{2'b01} & \textbf{2'b10} & \textbf{2'b11} \\ \hline
        & & \textbf{RGM BRAM 0} & \textbf{RGM BRAM 1} & \textbf{RGM BRAM 2} & \textbf{RGM BRAM 3}  \\ \hline
        \multirow{2}{*}&{\textbf{RESET}}  & \cellcolor{yellow}Wipe & \cellcolor{yellow}Wipe & \cellcolor{yellow}Wipe &  \\ \hline
        \textbf{SLOT 0} & \textbf{REG OPR} & \cellcolor{green}Write & & &  \\ \hline
        \textbf{SLOT 1} & \textbf{REG OPR} & \cellcolor{red!50}Read & \cellcolor{green}Write & & \cellcolor{yellow}Wipe \\ \hline
        \textbf{SLOT 2} & \textbf{REG OPR} & \cellcolor{yellow}Wipe & \cellcolor{red!50}Read & \cellcolor{green}Write & \\ \hline
        \textbf{SLOT 3} & \textbf{REG OPR} & & \cellcolor{yellow}Wipe & \cellcolor{red!50}Read & \cellcolor{green}Write \\ \hline
    \end{tabular}
    \label{tab:RGM_BRAM}
\end{table}
    Upon reset, Bank $0$ (BRAM $0$ and BRAM $1$) and Bank $1$ (BRAM $2$) are wiped sequentially. After this initialization, subsequent operations (wiping, reading, and writing) are performed in parallel across the banks, as illustrated in the Table \ref{tab:RGM_BRAM}.
    
    \item Input coefficient data arrangement: The precoding multiplier requires matrix to be formatted in a specific order, which differs from the way O-RAN delivers data to the memory module. Before writing to the Precoding Coefficient Memory (PCM), the input data is rearranged so that all even coefficients of the first column are read first, followed by the odd coefficients of the same column.
    \item Precoder Coefficient Memory (PCM) input: This stage is responsible for selecting the appropriate ping-pong memory buffer for storing the PCM data. %The address for storing the coefficients in the PCM is calculated based on the forwarded IAN and layer\textunderscore ID, as follows in the equation Eq.~\ref{eq:even}, and Eq.~\ref{eq:odd}: 
    
   % For even coefficient indices:
      %  \begin{equation} \label{eq:even}
        % IAN \times (\text{no. of data packets per column}) + (\text{layer\_ID}).
       % \end{equation}
   % For odd coefficient indices:
       %\begin{equation} \label{eq:odd}
        %IAN \times (\text{no. of data packets per %column}) + 8 + (\text{layer\_ID}).
       %\end{equation}
    \item RGM Sequence Lookup: Upon a read trigger, this stage retrieves the sequence of precoder matrices and initiates their reading from the PCM for transmission. Additionally, it communicates the number of times each matrix should be reused and indicates whether to load a new matrix through the mult\textunderscore cfg port.

    %\begin{figure}[ht]
    %\centering
    %\includegraphics[width=0.5\textwidth]{mat_coeff.png}
    %\caption{PCM TX Output format}
    %\label{fig: PCM TX Output format}
    %\end{figure}

\begin{table}[h!]
\centering
\caption{TX precoder coefficient matrix output format}
\label{tab:TX Precoder coefficient matrix output format}
\renewcommand{\arraystretch}{1.0}
\setlength{\tabcolsep}{3pt}
\begin{tabular}{|c|c|c|c|c|c|c|c|}
\hline
\rowcolor{orange!50} r0c0 & r0c1 & r0c2 & r0c3 & r0c4 & r0c5 & r0c6 & r0c7 \\ \hline
\rowcolor{orange!50} r1c0 & r1c1 & r1c2 & r1c3 & r1c4 & r1c5 & r1c6 & r1c7 \\ \hline

\rowcolor{yellow!50} r2c0 & r2c1 & r2c2 & r2c3 & r2c4 & r2c5 & r2c6 & r2c7 \\ \hline
\rowcolor{yellow!50} r3c0 & r3c1 & r3c2 & r3c3 & r3c4 & r3c5 & r3c6 & r3c7 \\ \hline

\rowcolor{pink!50} r4c0 & r4c1 & r4c2 & r4c3 & r4c4 & r4c5 & r4c6 & r4c7 \\ \hline
\rowcolor{pink!50} r5c0 & r5c1 & r5c2 & r5c3 & r5c4 & r5c5 & r5c6 & r5c7 \\ \hline

\rowcolor{red!50} r6c0 & r6c1 & r6c2 & r6c3 & r6c4 & r6c5 & r6c6 & r6c7 \\ \hline
\rowcolor{red!50} r7c0 & r7c1 & r7c2 & r7c3 & r7c4 & r7c5 & r7c6 & r7c7 \\ \hline

\rowcolor{brown!50} r8c0 & r8c1 & r8c2 & r8c3 & r8c4 & r8c5 & r8c6 & r8c7 \\ \hline
\rowcolor{brown!50} r9c0 & r9c1 & r9c2 & r9c3 & r9c4 & r9c5 & r9c6 & r9c7 \\ \hline

\rowcolor{gray!50} r10c0 & r10c1 & r10c2 & r10c3 & r10c4 & r10c5 & r10c6 & r10c7 \\ \hline
\rowcolor{gray!50} r11c0 & r11c1 & r11c2 & r11c3 & r11c4 & r11c5 & r11c6 & r11c7 \\ \hline

\rowcolor{purple!50} r12c0 & r12c1 & r12c2 & r12c3 & r12c4 & r12c5 & r12c6 & r12c7 \\ \hline
\rowcolor{purple!50} r13c0 & r13c1 & r13c2 & r13c3 & r13c4 & r13c5 & r13c6 & r13c7 \\ \hline

\rowcolor{violet!50} r14c0 & r14c1 & r14c2 & r14c3 & r14c4 & r14c5 & r14c6 & r14c7 \\ \hline
\rowcolor{violet!50} r15c0 & r15c1 & r15c2 & r15c3 & r15c4 & r15c5 & r15c6 & r15c7 \\ \hline
\end{tabular}

\end{table}
    \item PCM Output: This stage is responsible for reading the matrix at the specified (IAN) provided by the RGM sequencing block and transmitting it through the memory output ports-8 ports for a ($16\times8$) architecture, $16$ ports for a ($32\times8$) architecture, and $32$ ports for a ($64\times8$) architecture. Each memory output port reads two contiguous rows, one element at a time. Consequently, it takes $16$ clock cycles to read the entire matrix, regardless of the architecture. Each port's designated row is color-coded in Table  \ref{tab:TX Precoder coefficient matrix output format}, clearly indicating the row allocation per port.  
   \end{enumerate}
\subsection{Matrix multiplier module}
The matrix multiplier module performs large-scale matrix multiplications by multiplying a precoding matrix $H$ (of sizes ($16\times8$), ($32\times8$), or ($64\times8$), depending on the configuration) with the user data vectors ($X$) from the ORAN. The precoding matrix ($H$) represents the characteristics of the channel, while the data vector ($X$), arranged as an ($8\times1$) vector, enables spatial multiplexing across antenna layers.

The key components within the multiplier subsystem and their respective functions are as follows:
\begin{enumerate}
    \item User data broadcaster: This module distributes resource elements (RE) from the ORAN U-plane as $32$-bit coefficients. In an $8$-layer configuration, these are arranged into an ($8\times1$) vector ($X$) for multiplication with the channel matrix ($H$). The matrix ($H$) is divided into sub-matrices for easier handling. Each submatrix is divided into ($2\times8$) blocks. This structure enables efficient broadcast of user data in 1:2 or 1:4 increments, ensuring seamless alignment with ($H$) during multiplication.
    
    \item Dot product: The dot product module performs complex multiplications of the sub-matrices, such as a ($2\times8$) channel matrix with an ($8\times1$) user data vector, using an optimized Karatsuba approach\cite{complexpaz2023efficient}. Each element ($h$) in the channel matrix has a real part $a=Re(h)$ and an imaginary part $b = Im(h)$, while each data element ($x$) in the user matrix has a real part $c = Re(x)$ and an imaginary part $d = Im(x)$. The Karatsuba-based calculations (see Fig.\ref{fig:Dot Product of matrices}) proceed as follows:

    \begin{figure}[htbp]
            \centering
                \includegraphics[clip, trim=0.2cm 0.5cm 0.5cm 0.5cm, width=0.40 \textwidth]{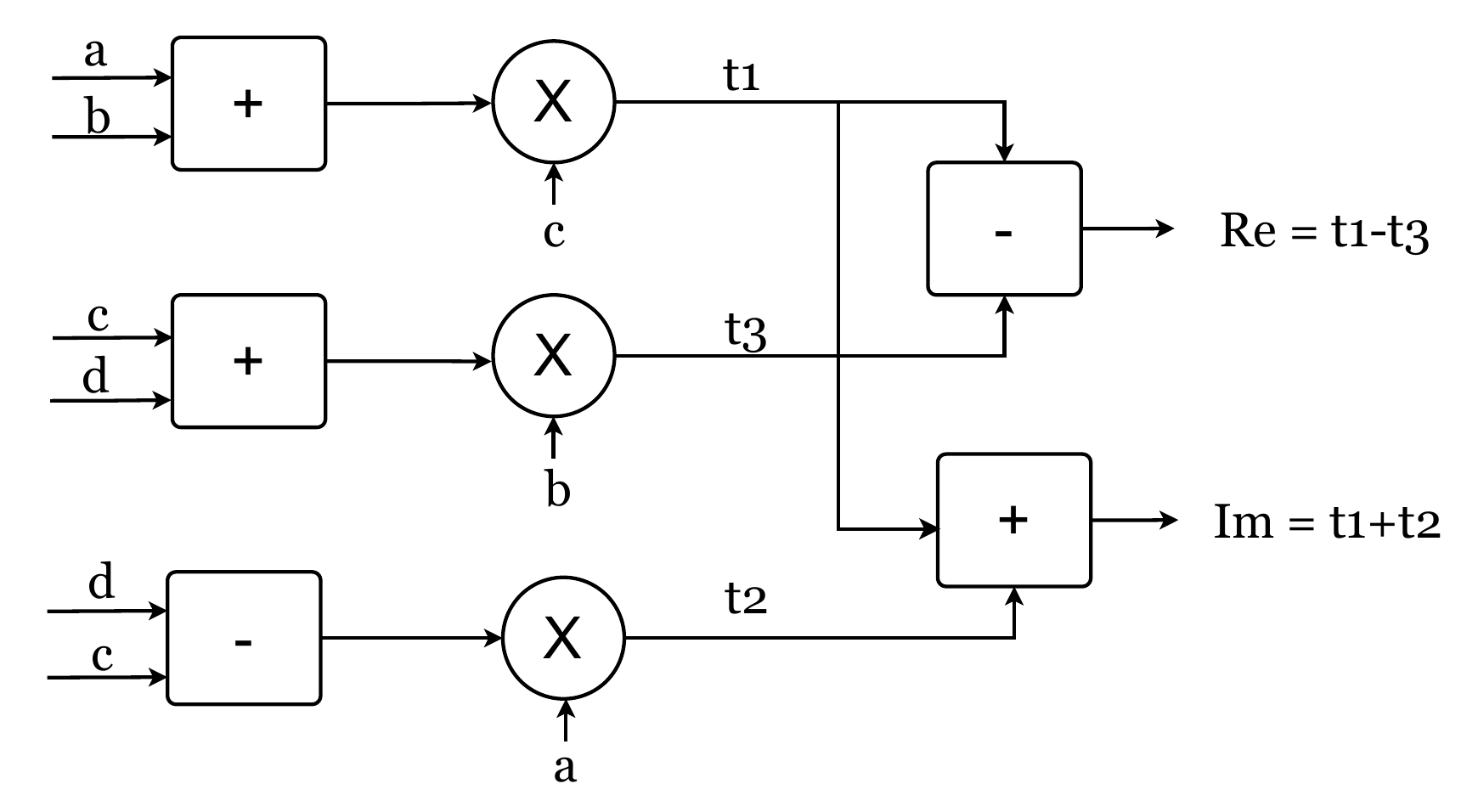}
            \caption{Dot product of matrices}
            \label{fig:Dot Product of matrices}
       \end{figure}

   \begin{enumerate}
        \item Compute three intermediate values:
        \begin{equation} \label{eq:t1}
            t_1 = c \times (a + b)
        \end{equation}
        \begin{equation}\label{eq:t2}
            t_2 = a \times (d - c)
        \end{equation}
        \begin{equation}\label{eq:t3}
            t_3 = b \times (c + d)
        \end{equation}
        
        \item Determine the real and imaginary outputs using equations Eq.~\ref{eq:t1}, Eq.~\ref{eq:t2}, and Eq.~\ref{eq:t3}:
        \begin{equation}
            \text{Re} = t_1 - t_3
        \end{equation}
        \begin{equation}
            \text{Im} = t_1 + t_2
        \end{equation}
\end{enumerate}

    For example, the multiplication of ($64\times8$)  is split into four ($16\times8$) multiplier modules. Each ($16\times8$) unit computes eight dot products between ($2\times8$) and ($8\times1$) matrices, with each dot product resulting in a  ($2\times1$) matrix output.

    \item Combiner/adder: This module aggregates the partial results generated from multiple dot products to form the final output as depicted in Fig.\ref{fig:Combiner/Adder}. 
    
   % \begin{figure}[ht]
    %\centering
    %\includegraphics[width=0.5\textwidth]{combiner.png}
    %\caption{Combiner/Adder}
    %\label{fig: Combiner/Adder}
    %\end{figure}

     \begin{figure}[htbp]
            \centering
                \includegraphics[clip, trim=0.2cm 0.5cm 0.5cm 0.5cm, width=0.45 \textwidth]{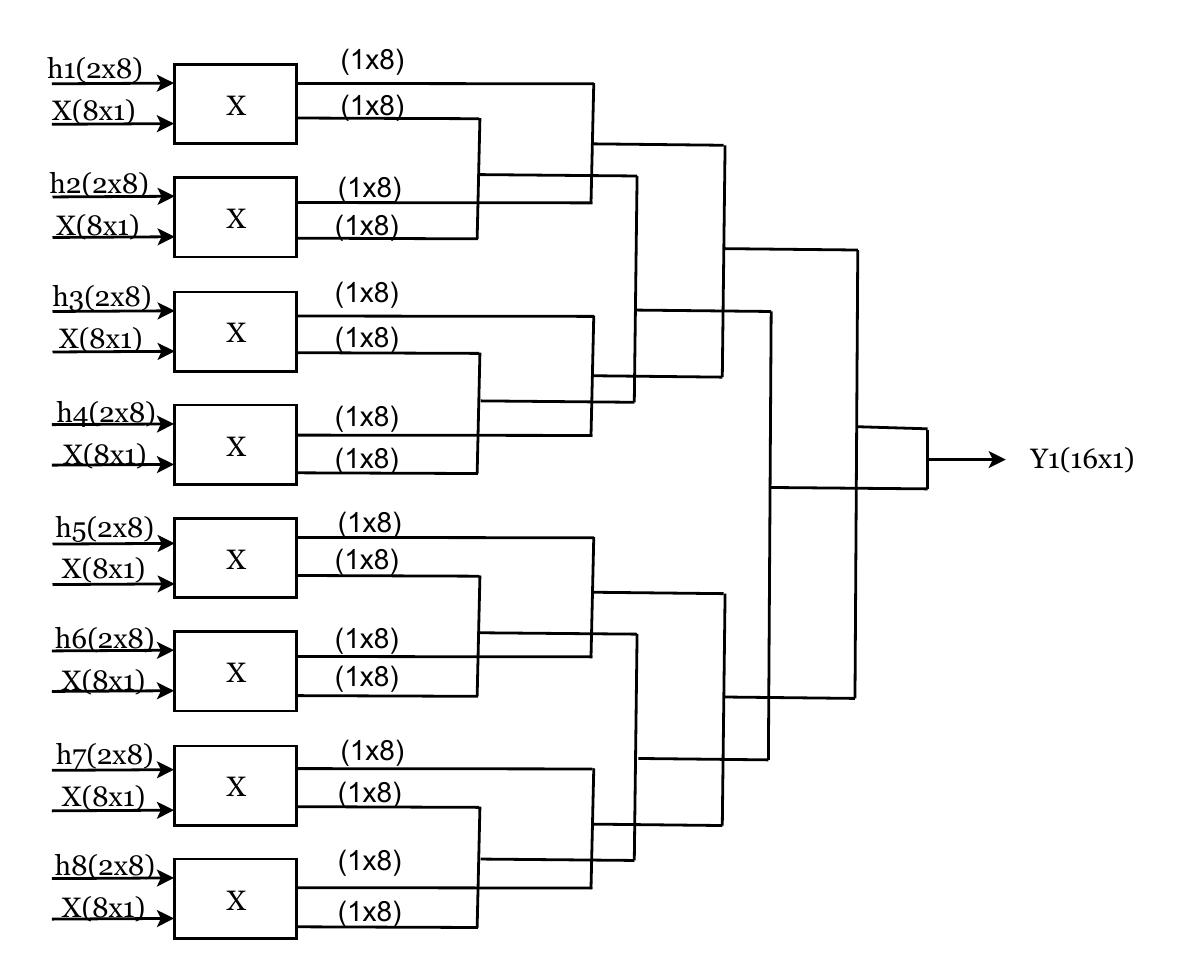}
            \caption{Combiner/Adder}
            \label{fig:Combiner/Adder}
       \end{figure}
    It takes the individual ($2\times1$) outputs from each dot product in the matrix multiplier and combines them. The combined output will be transmitted as ($Y$) in sizes of ($16\times1$), ($32\times1$) or ($64\times1$).

\end{enumerate}

\subsection{Precoder Tx to RX converter}
   The Tx and Rx precoders differ in the order of the matrix elements sent to the multiplier. While the Tx precoder transmits data in a specific sequence. In contrast, the Rx precoder memory uses a transposed version of the matrix, requiring a different data arrangement. This difference is clearly illustrated in the color-coded format shown in Table \ref{tab:RX PCM format}.
    
\begin{figure}[htbp]
    \centering
        \includegraphics[clip, trim=0.2cm 11cm 0.5cm 7cm, width=0.50\textwidth]{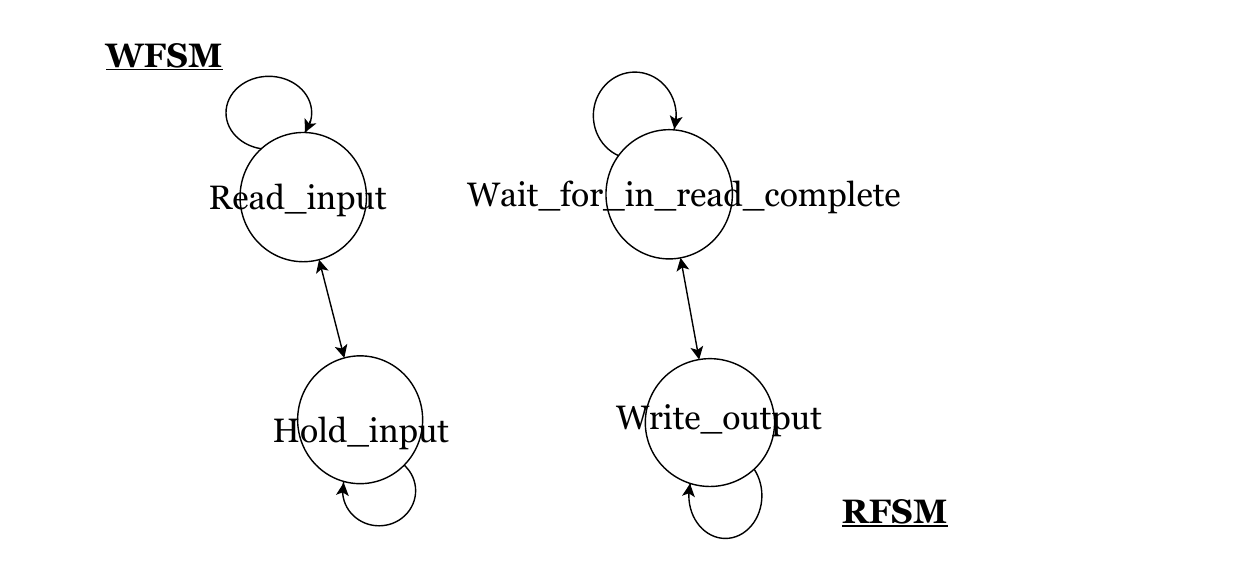}
    \caption{Read and Write FSM state diagram}
    \label{fig:State diagram}
\end{figure}

\begin{table}[h!]
\centering
\caption{RX precoder coefficient matrix output format}
\label{tab:RX PCM format}
\setlength{\tabcolsep}{0.0pt}

\begin{tabular}{|c|c|c|c|c|c|c|c|c|c|c|c|c|c|c|c|}
\hline
\rowcolor{orange!80} r0c0 & r0c1 & r0c2 & r0c3 & r0c4 & r0c5 & r0c6 & r0c7 & \cellcolor{orange!60}r0c8 & \cellcolor{orange!60}r0c9 & \cellcolor{orange!60}r0c10 & \cellcolor{orange!60}r0c11 & \cellcolor{orange!60}r0c12 & \cellcolor{orange!60}r0c13 & \cellcolor{orange!60}r0c14 & \cellcolor{orange!60}r0c15 \\ \hline
\rowcolor{orange!80} r1c0 & r1c1 & r1c2 & r1c3 & r1c4 & r1c5 & r1c6 & r1c7 & \cellcolor{orange!60}r1c8 & \cellcolor{orange!60}r1c9 & \cellcolor{orange!60}r1c10 & \cellcolor{orange!60}r1c11 & \cellcolor{orange!60}r1c12 & \cellcolor{orange!60}r1c13 & \cellcolor{orange!60}r1c14 & \cellcolor{orange!60}r1c15 \\ \hline
\rowcolor{yellow!60} r2c0 & r2c1 & r2c2 & r2c3 & r2c4 & r2c5 & r2c6 & r2c7 & \cellcolor{red!40}r2c8 & \cellcolor{red!40}r2c9 & \cellcolor{red!40}r2c10 & \cellcolor{red!40}r2c11 & \cellcolor{red!40}r2c12 & \cellcolor{red!40}r2c13 & \cellcolor{red!40}r2c14 & \cellcolor{red!40}r2c15 \\ \hline
\rowcolor{yellow!60} r3c0 & r3c1 & r3c2 & r3c3 & r3c4 & r3c5 & r3c6 & r3c7 & \cellcolor{red!40}r3c8 & \cellcolor{red!40}r3c9 & \cellcolor{red!40}r3c10 & \cellcolor{red!40}r3c11 & \cellcolor{red!40}r3c12 & \cellcolor{red!40}r3c13 & \cellcolor{red!40}r3c14 & \cellcolor{red!40}r3c15 \\ \hline
\rowcolor{purple!50} r4c0 & r4c1 & r4c2 & r4c3 & r4c4 & r4c5 & r4c6 & r4c7 & \cellcolor{brown!50}r4c8 & \cellcolor{brown!50}r4c9 & \cellcolor{brown!50}r4c10 & \cellcolor{brown!50}r4c11 & \cellcolor{brown!50}r4c12 & \cellcolor{brown!50}r4c13 & \cellcolor{brown!50}r4c14 & \cellcolor{brown!50}r4c15 \\ \hline
\rowcolor{purple!50} r5c0 & r5c1 & r5c2 & r5c3 & r5c4 & r5c5 & r5c6 & r5c7 & \cellcolor{brown!50}r5c8 & \cellcolor{brown!50}r5c9 & \cellcolor{brown!50}r5c10 & \cellcolor{brown!50}r5c11 & \cellcolor{brown!50}r5c12 & \cellcolor{brown!50}r5c13 & \cellcolor{brown!50}r5c14 & \cellcolor{brown!50}r5c15 \\ \hline
\rowcolor{blue!20} r6c0 & r6c1 & r6c2 & r6c3 & r6c4 & r6c5 & r6c6 & r6c7 & \cellcolor{green!10}r6c8 & \cellcolor{green!10}r6c9 & \cellcolor{green!10}r6c10 & \cellcolor{green!10}r6c11 & \cellcolor{green!10}r6c12 & \cellcolor{green!10}r6c13 & \cellcolor{green!10}r6c14 & \cellcolor{green!10}r6c15 \\ \hline
\rowcolor{blue!20} r7c0 & r7c1 & r7c2 & r7c3 & r7c4 & r7c5 & r7c6 & r7c7 & \cellcolor{green!10}r7c8 & \cellcolor{green!10}r7c9 & \cellcolor{green!10}r7c10 & \cellcolor{green!10}r7c11 & \cellcolor{green!10}r7c12 & \cellcolor{green!10}r7c13 & \cellcolor{green!10}r7c14 & \cellcolor{green!10}r7c15 \\ \hline
\end{tabular}

\end{table}
   This reordering is necessary due to the ($2\times 8$) dot product units in the multiplier module, which can only process $8$ columns of dot products at a time. The converter module arranges the transposed matrix to meet this processing constraint, allowing the main module of the Tx precoder to be easily used for Rx operations.
   The state diagram shown in Fig.\ref{fig:State diagram} describes the read and write finite state machine (FSM) of the Tx to RX converter module. 

     The function of the write FSM is as follows: \begin{enumerate}
        \item Read\textunderscore input: Load the incoming beamforming matrix in the TX sequence.
        \item Hold\textunderscore output: Once the input read is completed, hold this matrix until the output side can make a copy of it.
    \end{enumerate}

    The function of the read FSM is as follows: 
    \begin{enumerate}
        \item wait\textunderscore for\textunderscore in\textunderscore read\textunderscore complete: Wait until the input read is completed and then make a copy of the read matrix.
        \item write\textunderscore output: write the copy in the RX output sequence format. 
    \end{enumerate}    
\section{Data flow}
    \subsection{Algorithmic description of the TX precoder module}
       \begin{figure}[htbp]
            \centering
                \includegraphics[clip, trim=0.2cm 0.5cm 0.5cm 0.5cm, width=0.480 \textwidth]{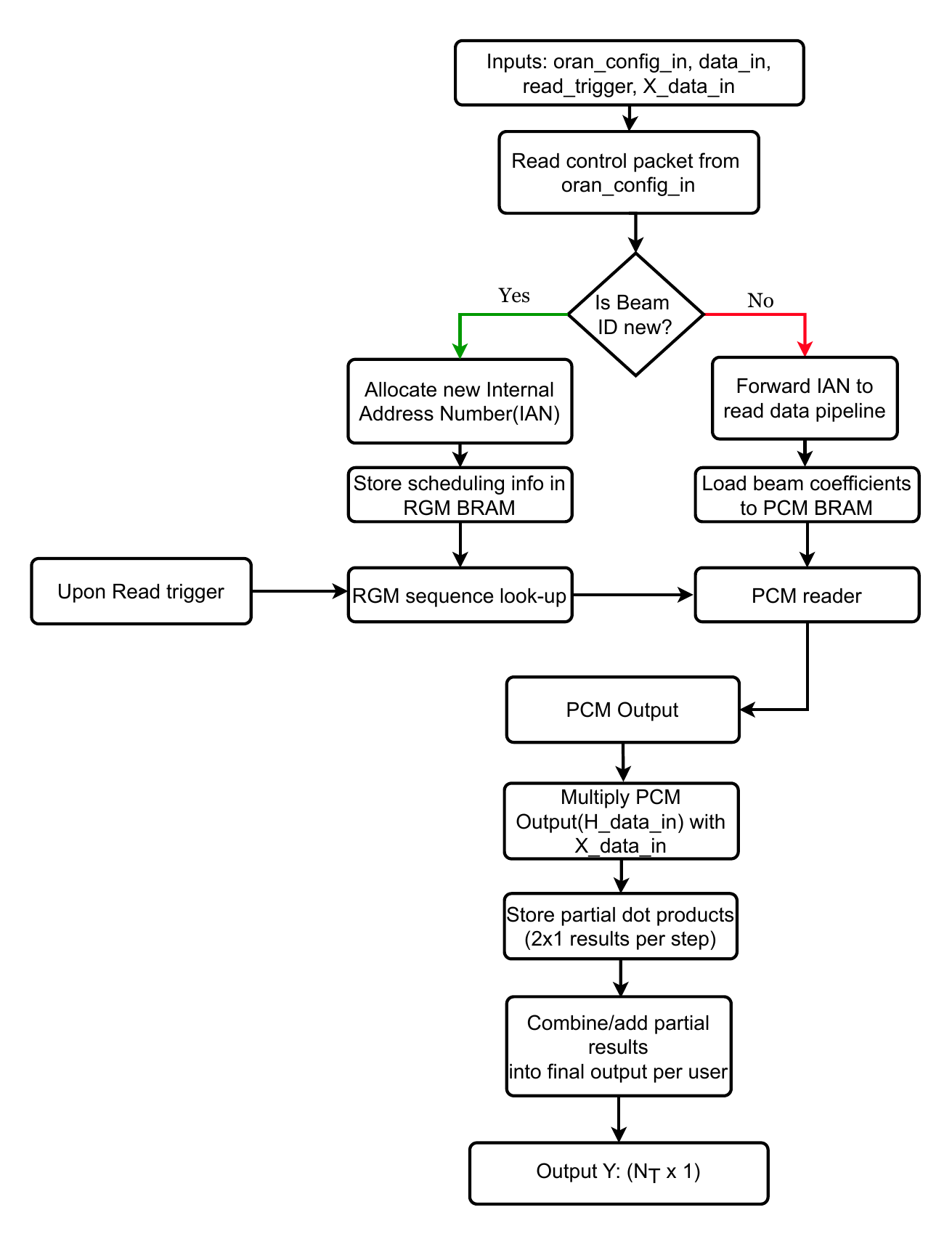}
            \caption{Algorithmic description of the TX precoder}
            \label{fig:Algorithmic description of the TX precoder}
       \end{figure}
The data flow in the TX precoder module is enumerated as follows (See Fig.\ref{fig:Algorithmic description of the TX precoder}):
    \begin{enumerate}
        \item Read a control packet from oran\textunderscore config\textunderscore in and find out if the beam id has appeared before or is this the first time this slot.
        \item If the beam id is new, allocate a new IAN for it, or if the beam is old, forward the allocated IAN to the read data pipeline.
        \item Read the beam coefficients for one layer using the data in port. Retrieve the IAN forwarded by the control pipeline. Store the data at the designated matrix address in the PCM BRAM buffer, following the order required for the multiplier.
        \item Meanwhile, the subsequent stages of the control pipeline store the scheduling information in the RGM BRAM buffer.
        \item Continue until a new control packet indicates a slot change. If both buffers are full, wait for one to free up; otherwise, alternate between writing and reading buffers.
        \item Upon receiving a read\_trigger, check the availability of the pipeline. If busy, acknowledge with a valid and ready signal and complete two reads before the next response. 
        \item When it is triggered, read the RGM to determine the sequence of matrices for the multiplier.
        \item Communicate the sequence and matrix usage count to the multiplier via mult\textunderscore cfg port and start reading the matrix from the PCM.
        \item The PCM data is then multiplied with X\textunderscore data\textunderscore in to perform the necessary beamforming operations.
        \item Partial dot products (($2\times1$) results per step) are temporarily stored as intermediate results.
        \item The stored partial results are combined or added to generate the final output for each user.
        \item The final output, represented as ($Y$), is produced as a vector with dimensions ($16\times1$), ($32\times1$) or ($64\times1$), depending on the configuration of the system.
        
    \end{enumerate}

\section{Testing and Evaluation}

\subsection{Simulation level}
      The precoder memory and multiplier modules were first tested individually using MATLAB-generated reference data and RTL simulations on the Xilinx Vivado platform. The functionality of each module was checked by comparing its output with the MATLAB results, to ensure that the modules performed as expected. To have an efficient data transfer towards and from the module, we adopt an AXIS handshake protocol for the I/O data ports. Inputs were applied directly to each module, and outputs were verified individually. After successful standalone testing, the modules were integrated into end-to-end transmitter and receiver chains to observe system-level performance. 
      
%To have an efficient data transfer towards and from the module, we adopt an AXIS handshake protocol for the I/O data ports.The precoder memory module has two clock domains, 312.5 MHz and 250 MHz, while the multiplier operates at 250 MHz and can scale up to 500 MHz. Key ports, such as data\_in and oran\_config\_in, load matrices and scheduling data from ORAN, while read\_trigger and mult\_cfg manage matrix loading and reuse. X\_in carries user data from ORAN, supporting up to $8$ layers, while H\_in receives coefficients from memory. The final beamforming outputs are then sent to the Digital Front End (DFE) through the Y\_out port. In simulation, each of these ports was tested to validate data flow, timing accuracy, and functionality. 

\subsection{Hardware Testing}
 After successful simulations, the precoder modules were integrated into the transmitter and receiver chain for hardware testing. Both wired and wireless configurations were tested using vector signal analyzer (VSA) and observed the system's performance and signal quality across different setups. Multiple stress tests were also performed to assess the robustness of the system, which shows that it can handle up to precoding of $48$ users within $500\mu s$. 

\section{Results} 
     Theoretical analysis based on the parameters shown in Table \ref{tab:Latency and Timing Parameters} shows that the read-out time of the memory module in the best case scenario is approximately $430\mu s$, which fits within the $500 \mu s$ slot boundary. This timing allows the precoder to operate at resource element-level granularity. The calculations below verify this performance. %supporting CSIR matrices without overshooting the slot duration. The calculations below verify this performance.
      
\begin{table}[h!]
\centering
\caption{Latency and Timing parameters}
\label{tab:Latency and Timing Parameters}
    \begin{tabular}{|l|c|}
        \hline
        \textbf{Parameter} & \textbf{Value} \\
        \hline
        Slot Boundary Time ($S_T$) & $500 \mu$s \\
        \hline
        Total No. of PRBs ($T_{\text{PRB}}$) & $273$ \\
        \hline
        $1$ RB  & $12$ RE\\
        \hline
        Total no. users ($T_u$) & $64$ \\
        \hline
        Latency to Load One Matrix ($T_{\text{load}}$) & $16$ clk cycles \\
        \hline
        Latency to Multiply One Matrix ($T_{\text{mult}}$)& $2 $ clk cycle \\
        \hline
        Time Period of Multiplier ($T_{\text{clk}}$) & $4$ ns \\
        \hline
        Max no. of H matrices ($n_H$) & $64$ \\
        \hline
        Max no. of user matrices ($n_X$) & $273 \times 14 \times 12 = 45864$ \\
        \hline
        Total number of multiplications ($N_{\text{mult}}$) & $\max(n(H), n(X)) = 45864$ \\
        \hline
        
\end{tabular}
\end{table}
\begin{table}
    \centering
    \caption{Timing analysis for different system configurations}
    \label{tab:Timing Analysis for Different Systems}
    \renewcommand{\arraystretch}{1.2}
    \begin{tabular}{|p{1.45cm}|p{1.9cm}|p{1.8cm}|p{2.0cm}|}
    \hline
     & \textbf{Setup Time (ns)} & \textbf{Hold Time (ns)} & \textbf{Pulse Width (ns)} \\ \hline
    16x8 system    & 1.133            & 0.026        &    1.105    \\ \hline          
    32x8 system      & 1.219            & 0.027        &    1.105     \\ \hline          
    64x8 system        & 1.831            & 0.045        &    1.105      \\ \hline       
    \end{tabular}
    \end{table}

\begin{table*}[h!]
\centering
\caption{Hardware resource utilization for different system configurations}
\label{tab:Resource Utilization for Different Systems}
\renewcommand{\arraystretch}{1.5}
\begin{tabular}{|p{1.0cm}|p{1.2cm}|p{1.2cm}|p{1.65cm}|p{1.2cm}|p{1.2cm}|p{1.65cm}|p{1.2cm}|p{1.2cm}|p{1.65cm}|}
\hline
\multirow{2}{1.5cm}{\textbf{Resource}} & \multicolumn{3}{c|}{\textbf{16x8 System}} & \multicolumn{3}{c|}{\textbf{32x8 System}} & \multicolumn{3}{c|}{\textbf{64x8 System}} \\ \cline{2-10} 
                          & \textbf{Utilization} & \textbf{Available} & \textbf{Utilization \%} & \textbf{Utilization} & \textbf{Available} & \textbf{Utilization \%} & \textbf{Utilization} & \textbf{Available} & \textbf{Utilization \%} \\ \hline
LUT                       & 2412        & 522720     & 0.46\%        & 3477       & 522720     & 0.67\%       & 5635       & 522720     & 1.08\%       \\ \hline
LUTRAM                    & 48         & 161280     & 0.03\%        & 48        & 161280     & 0.03\%       & 48        & 161280     & 0.03\%       \\ \hline
FF                        & 2739       & 1045440    & 0.26\%        & 3928       & 1045440    & 0.38\%       & 6145       & 1045440    & 0.59\%       \\ \hline
BRAM                      & 22.50          & 984       & 2.284\%       & 38.50          & 984      & 3.91\%       & 70.50         & 984       & 7.16\%       \\ \hline
DSP                       & 192          & 1968       & 9.75\%       & 384          & 1968       & 19.5\%       & 768         & 1968      & 39.00\%       \\ \hline
%IO                        & 2          & 308       & 0.65\%       & 2        & 308       & 0.65\%       & 2         & 308       & 0.65\%       \\ \hline
%BUFG                      & 1           & 940        & 0.11\%       & 1         & 940        & 0.11\%       & 1          & 940        & 0.11\%       \\ \hline

\end{tabular}
\end{table*}
      Considering there are 64 matrices across all 14 symbols, the optimal latency achievable is calculated using Eq.\ref{eq:matrix_ineq} and Eq.\ref{eq:user_eqn}.
          For each matrix denoted by $n_i$, ($i\;=\;1,2,...64$), the PRBs follow the inequality given by Eq.\ref{eq:matrix_ineq}.
          \begin{equation}\label{eq:matrix_ineq}
            (n_1 + n_2 + \dots + n_{64}) \leq T_{\text{PRB}}
         \end{equation}
          The matrix corresponding to the any user $n_k$ is calculated using the formula given by Eq.\ref{eq:user_eqn}.
          \begin{equation}\label{eq:user_eqn}
            \left(\frac{1}{T_{\text{clk}}} + T_{\text{mult}} \times 1 \, \text{RB}\right) \times \sum_{i=1}^{k} n_i + \left(T_{\text{load}} \times T_{\text{PRB}}\right) \; \text{clk cycles}
          \end{equation}  
     Using Eq.\ref{eq:matrix_ineq} and Eq.\ref{eq:user_eqn} the upper bound per symbol is determined to be $7644.25$ clk cycles, which corresponds to $107030$ clk cycles per slot, for a total of $14$ symbols. Therefore, the time required per slot for the multiplier operation is calculated as $430\mu s$ , which is within the $500\mu s$ slot duration.
   
     The analysis demonstrates that the maximum multiplier count, $N_{\text{mult}}$, occurs under a worst-case scenario where users are allocated alternately on the resource grid. In such a case, every alternating resource element is assigned to a new user, requiring a new matrix multiplication and hence loading of a new matrix in to the memory. However, if the user allocation pattern is different, the multiplier count can be lower, which may reduce the overall latency. 

    The timing analysis for the ($16\times8$), ($32\times8$), and ($64\times8$) configurations, presented in Table \ref{tab:Timing Analysis for Different Systems}, demonstrates that the setup time, hold time, and pulse width for each system meet the necessary timing constraints. Specifically, synchronization and latency checks confirm reliable operation with the multiplier operating at 250 MHz and memory access at 312.5 MHz. This ensures that the system works efficiently within the required timing limits.
    
    Table \ref{tab:Resource Utilization for Different Systems} provides an overview of the resource utilization across key hardware components (LUTs, FFs, BRAMs, and DSP blocks) for the ($16\times8$), ($32\times8$), and ($64\times8$) configurations. For the ($16\times8$) system, the utilization of DSPs, LUTs, FFs, and BRAMs is $9.75\%$, $0.46\%$, $0.26\%$, and $2.28\%$, respectively. This suggests that the ($16\times8$) configuration utilizes resources efficiently with relatively low consumption, indicating a well-balanced design. In the ($32\times8$) configuration, DSP utilization increases to $19.5$\%, and LUTs, FFs, and BRAMs consume $0.67$\%, $0.38$\%, and $3.91$\%, respectively. The higher DSP utilization in this configuration is expected due to the increase in processing demand. Finally, the ($64\times8$) system shows a significant increase in DSP usage, reaching $39$\%, while LUTs, FFs, and BRAMs account for $1.08$\%, $0.59$\%, and $7.16$\%, respectively. Despite the increase in resource consumption, particularly in DSPs and BRAMs, the system still operates within the capacity of the hardware platform xczu19eg-ffvd-1760-3-e (Zynq ultrascale + MPSoC family), confirming that the design scales efficiently while adhering to resource constraints. This analysis demonstrates that the system remains optimized for performance in varying configurations, operating efficiently without exceeding hardware limits. With $4$ active layers, the throughput achieved for both ($16\times8$) and ($32\times8$) channel configurations is $1.2 Gbps$, confirming the system’s effective use of available resources.

\section{Conclusion}
This paper presented an optimized precoder design for 5G NR, demonstrating efficient use of FPGA resources. With a throughput of $1.2 Gbps$ in both ($16\times8$) and ($32\times8$) configurations, the system achieves optimal performance using only four active layers. The design meets strict latency and resource constraints, confirming its feasibility for real-time 5G deployment. The flexibility across different configurations highlights the precoder’s robustness for future 5G and beyond applications.

% Furthermore, the designed physical layer is examined for the incorporation of Machine Learning (ML) models, as discussed in ~\cite{yerrapragada2023machine}.

\section*{Acknowledgment}
The authors extend their gratitude to the Department of Telecommunications (DOT), India for funding the 5G Testbed project.

% \bibliographystyle{IEEEtran}
% argument is your BibTeX string definitions and bibliography database(s)
% \bibliography{main}

% Generated by IEEEtran.bst, version: 1.14 (2015/08/26)

\end{document}